\documentclass[11pt]{article}
\usepackage{fullpage}
\usepackage{amssymb}

\newtheorem{theorem}{Theorem}[section]

\newtheorem{definition}[theorem]{Definition}

\newtheorem{claim}[theorem]{Claim}
\newtheorem{lemma}[theorem]{Lemma}

\newcommand{\qedsymb}{\hfill{\rule{2mm}{2mm}}}

\def\calf{{\cal F}}

\newcommand\set[2]{\left\{ #1 \left|\; #2 \right. \right\}}
\newcommand\sett[1]{\left\{ #1 \right\}}
\newcommand\defeq{\stackrel{def}{=}}
\newcommand\card[1]{\left| #1 \right|}

\newcommand\dbf[2]{\pi_{#1\to #2}}

\newcommand{\eps}{\varepsilon}
\newcommand{\ekvc}{E$k$-Vertex-Cover}
\renewcommand{\epsilon}{\varepsilon}

\newcommand{\blackslug}{\hbox{\hskip 1pt \vrule width 6pt height 5pt
depth 1.5pt \hskip 1pt}}
\newcommand{\QED}{\quad\blackslug\lower 8.5pt\null}
\newenvironment{proof}{\vspace{-6mm}\paragraph{Proof:}}{\hfill \QED \smallbreak}

\newcommand\remove[1]{{}}

\begin{document}

\title{A New Multilayered PCP and the Hardness of Hypergraph Vertex Cover}

\author{
Irit Dinur \footnote{NEC Research Institute, Princeton, NJ.
E-Mail: {\tt iritd@research.nj.nec.com. }}%
\and Venkatesan Guruswami\thanks{Department of Computer Science,
University of Washington, Seattle, WA 98195. E-Mail: {\tt
venkat@cs.washington.edu.}
Part of this work was done while the author was at UC Berkeley as a
Miller Research Fellow.} \and Subhash Khot\thanks{Department of
Computer Science, Princeton
University, Princeton, NJ 08544. E-Mail: {\tt khot@cs.princeton.edu}}%
\and Oded Regev \footnote{Institute for Advanced Study, Princeton,
NJ. E-Mail: {\tt odedr@ias.edu}. Research supported by NSF grant
CCR-9987845.}}

\maketitle
\begin{abstract}
Given a $k$-uniform hyper-graph, the E$k$-Vertex-Cover problem is
to find the smallest subset of vertices that intersects every
hyper-edge. We present a new multilayered PCP construction that
extends the Raz verifier. This enables us to prove that
E$k$-Vertex-Cover is NP-hard to approximate within factor
$(k-1-\epsilon)$ for any $k \geq 3$ and any $\epsilon>0$. The
result is essentially tight as this problem can be easily
approximated within factor $k$. Our construction makes use of the
biased Long-Code and is analyzed using combinatorial properties of
$s$-wise $t$-intersecting families of subsets.
\end{abstract}


\section{Introduction}

A $k$-uniform hypergraph $H=(V,E)$ consists of a set of vertices
$V$ and a collection $E$ of $k$-element subsets of $V$ called
hyperedges. A {\em vertex cover} of $H$ is a subset $S \subseteq
V$ such that every hyperedge in $E$ intersects $S$, i.e., $e \cap
S \neq \emptyset$ for each $e \in E$. An {\em independent set} in
$G$ is a subset whose complement is a vertex cover, or in other
words a subset of vertices that contains no hyperedge entirely
within it. The E$k$-Vertex-Cover problem is the problem of finding
a minimum size vertex cover in a $k$-uniform hypergraph. This
problem is alternatively called the minimum hitting set problem
with sets of size $k$ (and is equivalent to the set cover problem
where each element of the universe occurs in exactly $k$ sets).

The \ekvc\ problem is a fundamental NP-hard optimization problem
which arises in numerous settings. For $k=2$, it is just the
famous vertex cover problem on graphs. Owing to its NP-hardness,
one is interested in how well it can be approximated in polynomial
time. A very simple algorithm that is invariably taught in a
typical undergraduate algorithms class is the following: greedily
pick a maximal set of pairwise disjoint hyperedges and then
include all vertices in the chosen hyperedges in the vertex cover.
It is easy to show that this gives a factor $k$ approximation
algorithm for \ekvc. State of the art techniques yield only a tiny
improvement, achieving a $k-o(1)$ approximation ratio
\cite{Halperin}. This raises the question whether achieving an
approximation factor of $k-\epsilon$ for any constant $\epsilon>0$
could be NP-hard. \remove{This question is one of the central open
questions in the field of approximation algorithms (for instance,
it figures as the second entry in the list of open problems in the
final chapter of Vazirani's recent book~\cite{vijay}).}

In this paper, we prove a nearly tight hardness result for \ekvc.
Specifically, we prove that \ekvc\ is indeed NP-hard to
approximate within factor $(k-1-\eps)$ for any $\eps> 0$, thus
explaining why no efficient algorithm with performance guarantee
much better than $k$ has been found.

\vspace{-1ex}
\subsection*{Previous Hardness Results}

The vertex-cover problem on hypergraphs where the size of the
hyperedges is unbounded is nothing but the Set-Cover problem. For this
problem there is a $\ln n$ approximation
algorithm~\cite{lovasz-covers,johnsonapprox}, and a matching
$(1-o(1))\ln n$ hardness result due to Feige~\cite{feigeset}. The
first explicit hardness result shown for \ekvc\ was due to
Trevisan~\cite{Trevisan-bounded-degree} who considered the
approximability of bounded degree instances of several
combinatorial problems, and specifically showed an
inapproximability factor of $k^{1/19}$ for \ekvc.
Holmerin~\cite{Holmerin42} showed that E$4$-Vertex-Cover is
NP-hard to approximate within $(2-\epsilon)$. Independently,
Goldreich~\cite{Goldreich-hvc} showed a direct
`FGLSS'-type~\cite{FGLSS} reduction (involving no use of the
long-code, a crucial component in most recent PCP constructions)
attaining a hardness factor of $(2-\epsilon)$ for
E$k$-Vertex-Cover for some constant $k$. Later, Holmerin~\cite{Holmerin2}
showed that E$k$-Vertex-Cover is NP-hard to
approximate within a factor of $k^{1-\epsilon}$, and also that it
is NP-hard to approximate E$3$-Vertex-Cover within factor
$(3/2-\eps)$.

Somewhat surprisingly, more recently Dinur, Guruswami and Khot
gave a fairly {\em simple} proof of an $\alpha\cdot k$ hardness
result for \ekvc, (for some $\alpha > \frac 1 3$). The proof takes
a combinatorial view of Holmerin's construction and instead of
Fourier analysis uses some properties concerning intersecting
families of finite sets. The authors also give a more complicated
reduction that shows a factor $(k-3-\eps)$ hardness for \ekvc. The
crucial impetus for that work came from the recent result of Dinur
and Safra~\cite{DS-vc} on the hardness of approximating vertex
cover (on graphs), and as in \cite{DS-vc} the notion of biased
long codes and some extremal combinatorics relating to
intersecting families of sets play an important role. In addition
to ideas from \cite{DS-vc}, the factor $(k-3-\eps)$ hardness
result also exploits the notion of covering complexity introduced
by Guruswami, H\aa stad and Sudan~\cite{GHS}. Both the
$\alpha\cdot k$ and the $k-3-\epsilon$ results have not been
published (an ECCC manuscript exists, \cite{DGK}) since they have
been subsumed by the work presented herein.

\subsection*{Our result and techniques}

In this paper we improve upon all the above hardness results by
proving a factor $(k-1-\eps)$ inapproximability result for \ekvc.
Already for $k=3$, this is an improvement from $1.5-\epsilon$ to
$2-\epsilon$. Extending our result from $k-1-\epsilon$ to
$k-\epsilon$ appears highly non-trivial and in particular would
imply a factor $2-\epsilon$ hardness for vertex-cover on graphs, a
problem that is notoriously difficult. While our proof shares some
of the extremal combinatorics flavor of \cite{DS-vc} and
\cite{DGK}, it draws its strength mainly from a new multilayered
outer verifier system for NP languages. This multilayered system
is constructed using the Raz verifier~\cite{Raz} as a building
block.

The Raz verifier, which serves as the starting point or ``outer
verifier'' in most if not all recent hardness results, can be
described as follows. There are two sets of (non-Boolean)
variables $Y$ and $Z$, and for certain pairs of $y\in Y$ and $z\in
Z$, a constraint $\pi_{y \rightarrow z}$. The constraints are
projections, i.e., for each assignment to $y$ there exists exactly
one assignment to $z$ such that the constraint $\pi_{y \rightarrow
z}$ is satisfied. The goal is to find an assignment $A$ to the
variables so that a maximum number of constraints $\pi_{y
\rightarrow z}$ are satisfied, i.e., have the property $\pi_{y
\rightarrow z}(A(y)) = A(z)$. The PCP Theorem~\cite{AS,ALMSS}
along with the Parallel Repetition Theorem~\cite{Raz} imply that
for any $\epsilon>0$ it is NP-hard to distinguish between the case
where all the constraints can be satisfied and the case where no
more than a fraction $\epsilon$ of the constraints can be satisfied.

In \cite{DGK}, the $\alpha \cdot k$ hardness result is obtained by
replacing every $Y$ variable by a block of vertices
(representing its Long-Code). Hyperedges connect $y_1$-vertices to
$y_2$-vertices only if there is some $z\in Z$ such that $\pi_{y_1
\rightarrow z},\pi_{y_2 \rightarrow z}$ are constraints in the
system. This construction has an inherent symmetry between blocks
which deteriorates the projection property of the constraints,
limiting the hardness factor one can prove to at most $k/2$.

Another way of reducing the Raz verifier to \ekvc\ is by
maintaining the asymmetry between $Y$ and $Z$, introducing a block
of vertices for each variable in $Y$ and in $Z$ (representing
their Long-Code). Each constraint $\pi_{y \rightarrow z}$ can be
emulated by a set of hyperedges, where each hyperedge consists of
both $y$-vertices and $z$-vertices. The hyperedges can be chosen
so that if the initial PCP instance was satisfiable, then taking a
certain $1/k$ of the vertices in each block will be a
vertex-cover. However, this reduction has a basic `bipartiteness'
flaw: the underlying constraint graph, being bipartite with parts
$Y$ and $Z$, has a vertex cover of size at most one half of the
number of vertices. Taking all the vertices of, say, the $Z$
variables will be a vertex cover for the hypergraph regardless of
whether or not the initial PCP instance was satisfiable. This,
once again, limits the gap to no more than $k/2$.

We remark that this `bipartiteness' flaw naturally arises in other
settings as well. One example is approximate hypergraph coloring,
where indeed our multilayered PCP construction has been
successfully
used for showing hardness, see \cite{DRS-hypercol, Khot-hypercol}.\\

\noindent{\bf The Multilayered PCP.~} We overcome the $k/2$ limit
by presenting a new, multilayered PCP. In this construction we
maintain the projection property of the constraints that is a
strong feature of the Raz verifier, while overcoming the
`bipartiteness' flaw. In the usual Raz verifier we have two
`layers', the first containing the $Y$ variables and the second
containing the $Z$ variables. In the multilayered PCP, we have $l$
layers containing variables $X_1,X_2,\dots,X_l$ respectively.
Between every pair of layers $i_1$ and $i_2$, we have a set of
projection constraints that represent an instance of the Raz
verifier. In the multilayered PCP, it is NP-hard to distinguish
between (i) the case where there exists an assignment that
satisfies {\em all} the constraints (between every pair of
layers), and (ii) the case where for {\em every} $i_1,i_2$ it is
impossible to satisfy more than a fraction $\epsilon$ of the
constraints between $X_{i_1}$ and $X_{i_2}$.

In addition, we prove that the underlying constraint graph no
longer has the `bipartiteness' obstacle, i.e. it no longer has a
small vertex cover and hence a large independent set.
Indeed we show that the multilayered PCP has a
certain `weak-density' property: for any set containing an
$\epsilon$ fraction of the variables there are many constraints
between variables of the set. This guarantees that ``fake''
independent sets in the hypergraph (i.e., independent sets that
occur because there are no constraints between the variables of
the set) contain at most $\epsilon$ of the vertices. Hence, the
minimum vertex cover must contain vertices in almost all of the
blocks.

We mention that the PCP presented by Feige in \cite{feigeset} has
a few structural similarities with ours. Most notably, both have
more than two types of variables. However, while in our
construction the types are layered with decreasing domain sizes,
in Feige's construction the different types are all symmetric.
Furthermore, and more importantly, the constraints tested by the
verifier in Feige's construction are not projections while this is
a key feature of our multilayered PCP, crucially exploited
in our analysis.

We view the construction of the multilayered PCP as a central
contribution of our paper, and believe that it could be a powerful
tool to reduce from in other hardness of approximation results as
well. In fact, as mentioned above, our multilayered
construction has already been used in obtaining strong hardness
results for coloring $3$-uniform hypergraphs~\cite{DRS-hypercol,
Khot-hypercol} (namely the hardness of coloring a $2$-colorable
$3$-uniform hypergraph using an arbitrary constant number of
colors), a problem for which no non-trivial inapproximability
results are known using other techniques. We anticipate that this
new outer verifier will also find other applications besides the
ones in this paper and in \cite{DRS-hypercol, Khot-hypercol}.

\subsection*{The Biased Long-Code}
Our hypergraph construction relies on the Long-Code that was
introduced in \cite{BGS}, and more specifically, on the biased
Long-Code defined in \cite{DS-vc}. Thus, each PCP variable is
represented by a block of vertices, one for each `bit' of the
biased Long-Code. More specifically, in $x$'s block we have one
vertex for each subset of $R$, where $R$ is the set of assignments
for the variable $x$. However, rather than taking all vertices in
a block with equal weight, we attach weights to the vertices
according to the $p$-biased Long-Code. The weight of a subset $F$
is set to $p^{\card F}(1-p)^{\card{R\setminus F}}$, highlighting
subsets of cardinality $p\cdot \card R$. Thus we actually
construct a weighted hypergraph which can then be easily
translated, by appropriate duplication of vertices, to a
non-weighted one (see, e.g., \cite{DS-vc}).

The vertex cover in the hypergraph is shown to have relative size
of either $1-p$ in the good case or almost $1$ in the bad case.
Choosing large $p = 1 - \frac 1{k-1-\epsilon}$, yields the desired
gap of $\frac 1 {1-p} \approx {k-1-\epsilon}$ between the good and
bad cases. The reduction uses the following property: {\em a
family of subsets of a set $R$, where each subset has size $p\card
R$, either contains very few subsets, or it contains some $k-1$
subsets whose common intersection is very small. } We will later
show that this property holds for $p < 1-\frac 1{k-1}$ and
therefore we obtain a gap of $k-1-\epsilon$. As can be seen, this
property does not hold for $p
> 1-\frac 1{k-1}$ and therefore one cannot improve the
$k-1-\epsilon$ result by simply increasing $p$.

\vspace{-1ex}
\subsection*{Location of the gap} All our hardness results have the gap
between sizes of the vertex cover at the ``strongest'' location.
Specifically, to prove a factor $(k-1-\epsilon)$ hardness we show
that it is hard to distinguish between $k$-uniform hypergraphs
that have a vertex cover of weight $\frac{1}{k-1}+ \epsilon$ from
those whose minimum vertex cover has weight at least
$(1-\epsilon)$. This result is stronger than a gap of about
$(k-1)$ achieved, for example, between vertex covers of weight
$\frac 1{(k-1)^2}$ and $\frac 1{k-1}$. In fact, by adding dummy
vertices, our result implies that for any $c<1$ it is NP-hard to
distinguish between hypergraphs whose minimum vertex-cover has weight
at least $c$ from those which have a vertex-cover of weight at most
$\bigl(\frac{c}{k-1} + \eps\bigr)$. Put another way, our result
shows that for $k$-uniform hypergraphs, for $k \ge 3$, there is a
fixed $\alpha$ such that for arbitrarily small $\epsilon > 0$, it
is NP-hard to find an independent set consisting of a fraction
$\epsilon$ of the vertices even if the hypergraph is promised to
contain an independent set comprising a fraction $\alpha$ of
vertices. We remark that such a result is not known for graphs and
seems out of reach of current techniques. (The recent $1.36$
hardness result for vertex cover on graphs due to Dinur and
Safra~\cite{DS-vc}, for example, shows that it is NP-hard to
distinguish between cases when the graph has an independent set of
size $0.38\cdot n$ and when no independent set has more than $0.16
\cdot n$ vertices.)

\subsection*{Organization} We begin in Section~\ref{sec:combin} by
developing the machinery from extremal combinatorics concerning
intersecting families of sets that will play a crucial role in our
proof. In Section~\ref{sec:layeredPCP} we present the
multilayered PCP construction.  In
Section~\ref{sec:reduction} we present our reduction to a gap
version of \ekvc\ which allows us to prove a factor $(k-1-\eps)$
inapproximability result for this problem.

\vspace{-1ex}
\section{Intersecting Families}
\label{sec:combin} In this section we describe certain properties
of $s$-wise $t$-intersecting families. For a comprehensive survey,
see \cite{HandbookOfCombinatorics}. Denote
$[n]=\sett{0,1,\ldots,n-1}$ and  $2^{[n]} = \sett{ F \ | \ F
\subseteq [n]}$.
\begin{definition}
A family $\calf \subseteq 2^{[n]}$ is called {\em $s$-wise
$t$-intersecting} if for every $s$ sets  $F_1,\ldots,F_s \in
\calf$, we have $$ \ |F_1\cap \ldots \cap F_s| \ge t\,.$$
\end{definition}

We are interested in bounding the size  of such families, and for
this purpose it is useful to introduce the notion of a
left-shifted family. Performing an $(i,j)$-shift on a family
consists of replacing the element $j$ with the element $i$ in all
sets $F\in \calf$ such that $j\in F$, $i\notin F$ and $(F\setminus
\{j\}) \cup \{i\} \notin \calf$. A left-shifted family is a family
which is invariant with respect to $(i,j)$-shifts for any $1\le
i<j\le n$. For any family $\calf$, by iterating the $(i,j)$-shift
for all $1\le i<j\le n$ we eventually get a left-shifted family
which we denote by $S(\calf)$. The following simple lemma
summarizes the properties of the left-shift operation (see, e.g.,
\cite{HandbookOfCombinatorics}, p. 1298, Lemma 4.2):
\begin{lemma}\label{basic_left_shift}
For any family $\calf\subseteq 2^{[n]}$, there exists a one-to-one
and onto mapping $\tau$ from $\calf$ to $S(\calf)$ such that
$|F|=|\tau(F)|$ for every $F\in \calf$. In other words,
left-shifting a family maintains its size and the size of the sets
in the family. Moreover, if $\calf$ is an $s$-wise
$t$-intersecting family then so is $S(\calf)$. \qedsymb
\end{lemma}

The next lemma states that a subset $F$ in a left-shifted $s$-wise
$t$-intersecting family, cannot be `sparse' on all of its prefixes
$F \cap [t+js],\;\forall j\ge 0$.
\begin{lemma}[\cite{HandbookOfCombinatorics}, p. 1311, Lemma 8.3]
\label{property_left_shift} Let $\calf$ be a left-shifted $s$-wise
$t$-intersecting family. Then, for every $F\in \calf$, there
exists a $j\ge 0$ with $|F \cap [t+sj]|\ge t+(s-1)j$. \qedsymb
\end{lemma}


\begin{definition}\label{def:biasedmu} For a bias parameter $0 < p <
1$, and a ground set $R$, the weight of a set $F\subseteq R$ is
$$\mu_p^R(F)\defeq p^{|F|}\cdot(1-p)^{|R\setminus F|}$$
When $R$ is clear from the context we write $\mu_p$ for $\mu_p^R$.
The weight of a family $\calf \subseteq 2^{R}$ is $\mu_p(\calf) =
\sum_{F\in \calf} \mu_p(F)$.
\end{definition}
The weight of a subset is precisely the probability of obtaining
this subset when one picks every element in $R$ independently with
probability $p$.

\medskip
The following is the main lemma of this section. It shows that for
any $p < \frac {s-1}s$, a family of non-negligible $\mu_p$-weight
(i.e., $\mu_p(\calf) \ge \epsilon$) cannot be $s$-wise
$t$-intersecting for sufficiently large $t$.
\begin{lemma}\label{intersecting_lemma}
For any $\epsilon,s,p$ with $p<\frac{s-1}{s}$,  there exists a
$t=t(\epsilon,s,p)$ such that for any $s$-wise $t$-intersecting
family $\calf\subseteq 2^{[n]}$, $\mu_p(\calf) < \epsilon$.
\end{lemma}
\begin{proof}
The proof follows from Lemma \ref{property_left_shift}
(see \cite{HandbookOfCombinatorics}, p. 1311, Theorem 8.4). Let
$\calf$ be an $s$-wise $t$-intersecting family where $t$ will be
determined later. According to Lemma~\ref{basic_left_shift},
$S(\calf)$ is also $s$-wise $t$-intersecting and
$\mu_p(S(\calf))=\mu_p(\calf)$. By
Lemma~\ref{property_left_shift}, for every $F\in S(\calf)$, there
exists a $j\ge 0$ such that $|F \cap [t+sj]|\ge t+(s-1)j$.  We
can therefore bound $\mu_p(S(\calf))$ from above by the probability that
such a $j$ exists for a random set chosen according to the
distribution $\mu_p$. We now prove an upper bound on this
probability, which will give the desired bound on $\mu_p(S(\calf))$
and hence also on $\mu_p(\calf)$.

Let $\delta = \frac{s-1}{s}-p$. Then, for any $j\ge 0$, $\mbox{Pr}
[~|F \cap [t+sj]|\ge t+(s-1)j~]$ is at most
\[   \mbox{Pr}[~|F \cap [t+sj]| - p(t+sj) \ge \delta(t+sj)~] \le
e^{-2(t+sj)\delta^2} \ . \]
by the Chernoff bound~\cite{Chernoff}. Summing over all $j\ge 0$
we get:
$$
\mu_p(S(\calf)) \le \sum_{j\ge 0} e^{-2(t+sj)\delta^2} =
e^{-2t\delta^2}/(1-e^{-2s\delta^2})
$$
which is smaller than $\epsilon$ for large enough $t$.
\end{proof}

\section{The Multilayered PCP}
\label{sec:layeredPCP}

\subsection{Starting Point - The PCP Theorem and the
Parallel Repetition Theorem}

As is the case with many inapproximability results (e.g.,
\cite{BGS}, \cite{Has96}, \cite{Has97}, \cite{SaT}), we begin our
reduction from the Raz verifier described next. Let $\Psi$ be a
collection of two-variable constraints, where the variables are of
two types, denoted $Y$ and $Z$. Let $R_Y$ denote the range of the
$Y$-variables and $R_Z$ the range of the $Z$-variables, where
$\card{R_Z}\le\card{R_Y}$\footnote{Readers familiar with the Raz
verifier may prefer to think concretely of $R_Y = [7^u]$ and $R_Z
= [2^u]$ for some number $u$ of repetitions.}. Assume each
constraint $\pi\in\Psi$ depends on exactly one $y\in Y$ and one
$z\in Z$, furthermore, for every value $a_y\in R_Y$ assigned to
$y$ there is exactly one value $a_z\in R_Z$ to $z$ such that the
constraint $\pi$ is satisfied. Therefore, we can write each
constraint $\pi\in\Psi$ as a function from $R_Y$ to $R_Z$, and use
notation $\dbf y z:R_Y \to R_Z$. Furthermore, we assume that the
underlying constraint graph is bi-regular, i.e., every
$Y$-variable appears in the same number of constraints in $\Psi$,
and every $Z$-variable appears in the same number of constraints
in $\Psi$.

The following theorem follows by combining the PCP Theorem with
Raz's Parallel Repetition Theorem. The PCP given by this theorem
will be called the Raz's verifier henceforth.
\begin{theorem}
\label{thm:Raz-PCP}
~(PCP Theorem \cite{ALMSS,AS} +
Raz's Parallel Repetition Theorem \cite{Raz})~~
Let $\Psi$ be as above. There exists a universal constant
$\gamma>0$ such that for every (large enough) constant $\card
{R_Y}$ it is NP-hard to distinguish between the following two cases:

\begin{itemize}
\item {\bf Yes : } There is an assignment $A:Y \to R_Y$,
$A: Z\to R_Z$ such that all $\pi\in \Psi$ are satisfied by $A$,
i.e., $\forall \dbf y z \in \Psi, \; \dbf y z (A(y))=A(z)$.

\item {\bf No : } No assignment can satisfy more than a fraction
$\frac{1}{\card{R_Y}^\gamma}$ of the constraints in $\Psi$.
\qedsymb
\end{itemize}

\end{theorem}

As discussed in the introduction, a natural approach to build a
hypergraph from the PCP $\Psi$ is to have a block of vertices for
every variable $y$ or $z$ and define hyperedges of the hypergraph
so as to enforce the constraints $\dbf y z$. For every constraint
$\dbf y z$, there will be hyperedges containing vertices from the
block of $y$ and the block of $z$. However, this approach is
limited by the fact that the constraint graph underlying the PCP
has a small vertex cover. Since each hyperedge contains vertices
from both the $Y$ and $Z$ `sides', the subset of all vertices on
the $Y$ (resp. $Z$) `side', already covers all of the hyperedges
regardless of whether the initial PCP system was satisfiable or
not.\footnote{Adding hyperedges entirely within vertices on the
$Y$ and $Z$ sides cannot help either since we wish to ensure a
small vertex cover in the completeness case. Hence picking all
vertices on, say, the $Z$ side, together with the small vertex
cover that hits all edges entirely within the $Y$ side (such a
small cover must exist due to the completeness case) will again
give a vertex cover of weight close to $1/2$.}

This difficulty motivates our construction of a multilayered PCP
where we have many types of variables (rather than only $Y$ and
$Z$) and the resulting hypergraph is {\em multipartite}. The
multilayered PCP is able to maintain the properties of Theorem
\ref{thm:Raz-PCP} between {\em every} pair of layers. Moreover,
the underlying constraint graph has a special `weak-density'
property that roughly guarantees it will have only tiny
independent sets (thus any vertex cover for it must contain almost
all of the vertices).

\subsection{Layering the Variables}

Let $l,R>0$. Let us begin by defining an $l$-layered PCP. In an
$l$-layered PCP there are $l$ sets of variables denoted by
$X_1,\ldots,X_l$. The range of variables in $X_i$ is denoted
$R_i$, with $\card{R_i} = R^{O(l)}$. For every $1\le i < j\le l$
there is a set of constraints $\Phi_{ij}$ where each constraint
$\pi \in \Phi_{ij}$ depends on exactly one $x\in X_i$ and one
$x'\in X_j$. For any two variables we denote by $\dbf x {x'}$ the
constraint between them if such a constraint exists. Moreover, the
constraints in $\Phi_{ij}$ are projections from $x$ to $x'$, that
is, for every assignment to $x$ there is exactly one assignment to
$x'$ such that the constraint is satisfied.

In addition, as mentioned in the introduction, we would like to
show a certain `weak-density' property of our multilayered PCP:
\begin{definition}\label{def:weak-density}
An $l$-layered PCP is said to be {\em weakly-dense} if for
any $\delta>0$,  given $m\ge \lceil{\frac 2 \delta}\rceil$ layers
$i_1<\ldots<i_m$ and given any sets $S_j \subseteq X_{i_j}$ for
$j\in [m]$ such that $S_j \ge \delta |X_{i_j}|$,
 there always exist two sets $S_j$ and $S_{j'}$ such that the number of constraints
between them is at least a $\frac{\delta^2}{4}$ fraction of the
constraints between the layers $X_{i_j}$ and $X_{i_{j'}}$.
\end{definition}

\begin{theorem}\label{pcp_thm}
There exists a universal constant $\gamma>0$, such that for any
parameters $l,R$, there is a weakly-dense $l$-layered PCP
$\cup\Phi_{ij}$ such that  it is NP-hard to distinguish between
the following two cases:

\begin{itemize}
\item {\bf Yes : } There exists an assignment that satisfies all
the constraints.%
\item {\bf No : } For every $i<j$, not more than $1/R^\gamma$ of
the constraints in $\Phi_{ij}$ can be satisfied by an assignment.
\end{itemize}

\end{theorem}

\begin{proof}
Let $\Psi$ be a constraint-system as in Theorem~\ref{thm:Raz-PCP}.
We construct $\Phi = \cup\Phi_{ij}$ as follows. The variables
$X_i$ of layer $i\in [l]$ are the elements of the set $Z^i\times
Y^{l-i}$, i.e., all $l$-tuples where the first $i$ elements are
$Z$ variables and the last $l-i$ elements are $Y$ variables. The
variables in layer $i$ have assignments from the set
$R_i=(R_Z)^{i}\times (R_Y)^{l-i}$ corresponding to an assignment
to each variable of $\Psi$ in the $l$-tuple. It is easy to see
that $|R_i|\le R^{O(l)}$ for any $i\in [l]$ and that the total
number of variables is no more than $\card{\Psi}^{O(l)}$. For any
$1\le i<j\le l$ we define the constraints in $\Phi_{ij}$ as
follows. A constraint exists between a variable $x_i\in X_i$ and a
variable $x_j \in X_j$ if they contain the same $\Psi$ variables
in the first $i$ and the last $l-j$ elements of their $l$-tuples.
Moreover, for any $i< k\le j$ there should be a constraint in
$\Psi$ between $x_{i,k}$ and $x_{j,k}$. More formally, denoting
$x_i = (x_{i,1},...,x_{i,l})$ for $x_i \in X_i = Z^i\times
Y^{l-i}$,
\begin{eqnarray*}
\Phi_{ij} & = & \biggl\{ \pi_{x_i,x_j} ~\vline ~ x_i\in X_i,x_j\in X_j, \\
& & \qquad
  \forall k\in [l] \setminus \{i+1,\ldots,j\}, x_{i,k}=x_{j,k} \\
& & \qquad
  \forall k\in\{i+1,\ldots,j\}, \dbf{x_{i,k}}{x_{j,k}}\in \Psi
\biggr\}  \ .
\end{eqnarray*}
As promised, the constraints $\dbf{x_{i,k}}{x_{j,k}}$ are
projections. Given an assignment $a=(a_1,..,a_l)\in R_i$ to $x_i$,
we define the consistent assignment $b=(b_1,..,b_l)\in R_j$ to
$x_j$ as $b_k=\dbf{x_{i,k}}{x_{j,k}}(a_k)$ for
$k\in\{i+1,\ldots,j\}$ and $b_k=a_k$ for all other $k$.

The completeness of $\Phi$ follows easily from the completeness of
$\Psi$. That is, assume we are given an assignment $A:Y\cup Z
\rightarrow R_Y \cup R_Z$ that satisfies all the constraints of
$\Psi$. Then, the assignment $B:\bigcup X_i \rightarrow \bigcup
R_i$ defined by $B(x_1\ldots x_l)=(A(x_1) \ldots A(x_l))$ is a
satisfying assignment.

For the soundness part, assume that there exist two layers $i<j$ and an assignment $B$ that satisfies more than a
$1/R^\gamma$ fraction of the constraints in $\Phi_{ij}$. We partition $X_i$ into classes such that two variables in
$X_i$ are in the same class iff they are identical except possibly on coordinate $j$.  The variables in $X_j$ are also
partitioned according to coordinate $j$. Since more than $1/R^\gamma$ of the constraints in $\Phi_{ij}$ are satisfied,
it must be the case that there exist a class $x_{i,1},\ldots,x_{i,j-1},x_{i,j+1},\ldots,x_{i,l}$ in the partition of
$X_i$ and a class $x_{j,1},\ldots,x_{j,j-1},x_{j,j+1},\ldots,x_{j,l}$ in the partition of $X_j$ between which there
exist constraints and the fraction of satisfied constraints is more than $1/R^\gamma$. We define an assignment to
$\Psi$ as
\[ A(y) = (B(x_{i,1},\ldots,x_{i,j-1},y,x_{i,j+1},\ldots,x_{i,l}))_j \]
for $y\in Y$ and as
\[ A(z) = (B(x_{j,1},\ldots,x_{j,j-1},z,x_{j,j+1},\ldots,x_{j,l}))_j \]
for $z\in Z$. Notice that there is a one-to-one and onto
correspondence between the constraints in $\Psi$ and the constraints
between the two chosen classes in $\Phi$. Moreover, if the constraint
in $\Phi$ is satisfied, then the constraint in $\Psi$ is also
satisfied. Therefore, $A$ is an assignment to $\Psi$ that satisfies
more than $1/R^\gamma$ of the constraints.

To prove that this multilayered PCP is {\em weakly-dense}, we
recall the bi-regularity property mentioned above,
i.e., each variable $y\in Y$ appears in the same number of
constraints and also each $z\in Z$ appears in the same number of
constraints. Therefore, the distribution obtained by uniformly
choosing a variable $y\in Y$ and then uniformly choosing one of
the variables in $z\in Z$ with which it has a constraint is a uniform
distribution on $Z$.

Take any $m = \lceil{\frac 2 \delta}\rceil$ layers
$i_1<\ldots<i_m$ and sets $S_j \subseteq X_{i_j}$ for $j\in [m]$
such that $S_j \ge \delta |X_{i_j}|$. Consider a random walk
beginning from a uniformly chosen variable $x_1\in X_1$ and
proceeding to a variable $x_2\in X_2$ chosen uniformly among the
variables with which $x_1$ has a constraint. The random walk continues
in a similar way to a variable $x_3\in X_3$ chosen uniformly among
the variables with which $x_2$ has a constraint and so on up to a
variable in $X_l$. Denote by $E_j$ the indicator variable of the
event that the random walk hits an $S_j$ variable when in layer
$X_{i_j}$. From the uniformity of $\Psi$ it follows that for every
$j$, $\mbox{Pr}[E_j] \ge \delta$. Moreover, using the
inclusion-exclusion principle, we get:
\begin{eqnarray*}
 && 1 ~\ge
~\mbox{Pr}[\bigvee E_j]
 \ge \sum_j \mbox{Pr}[E_j]
 - \sum_{j<k} \mbox{Pr}[E_j \wedge E_k] \\
 && ~\ge ~\lceil \frac{2}{\delta} \rceil \cdot \delta - {m\choose 2}
\mbox{max}_{j<k} \mbox{Pr}[E_j \wedge E_k] \\
& &    ~\ge 2 - {m\choose 2} \mbox{max}_{j<k} \mbox{Pr}[E_j \wedge E_k]
\end{eqnarray*}
which implies
$$\mbox{max}_{j<k} \mbox{Pr}[E_j \wedge E_k]
 \ge 1/{m\choose 2} \ge \frac{\delta^2}{4} $$

Fix $j$ and $k$ such that $\mbox{Pr}[E_j \wedge E_k] \ge
\frac{\delta^2}{4}$ and consider a shorter random walk beginning from
a random variable in $X_{i_j}$ and proceeding to the next layer and so
on until hitting layer ${i_k}$. Since $E_j$ is uniform on $X_{i_j}$ we
still have that $\mbox{Pr}[E_j \wedge E_k]\ge \frac{\delta^2}{4}$
where the probability is taken over the random walks between $X_{i_j}$
and $X_{i_k}$. Also, notice that there is a one-to-one and onto
mapping from the set of all random walks between $X_{i_j}$ and
$X_{i_k}$ to the set $\Phi_{i_j,i_k}$. Therefore, at least a fraction
$\frac{\delta^2}{4}$ of the constraints between $X_{i_j}$ and
$X_{i_k}$ are between $S_j$ and $S_k$, which completes the proof of
the weak-density property.
\end{proof}

\section{The Hypergraph Construction}
\label{sec:reduction}
\begin{theorem}[Main Theorem]
For any $k \ge 3$ it is NP-hard to approximate the vertex-cover on
a $k$-uniform hypergraph  within any constant factor less than
$k-1$.
\end{theorem}
\begin{proof}
Fix $k \ge 3$ and arbitrarily small  $\epsilon>0$. Define $p =
1-\frac{1}{k-1-\epsilon}$. Let $\Phi$ be a PCP instance with
layers $X_1,\ldots,X_l$, as described in Theorem~\ref{pcp_thm},
with parameters $l=32\epsilon^{-2}$ and $R$ large enough to be
chosen later. We present a construction of a $k$-uniform
hypergraph $G = (V,E)$. We use the Long Code introduced by Bellare
et al. \cite{BGS}. A Long Code over domain $R$ has one bit for
every subset $v \subseteq R$. An encoding of element $x \in R$
assigns bit-value $1$ to the sets $v$ s.t. $x \in v$  and assigns
$0$ to the sets which do not contain $x$. In the following, the
bits in the Long Code will be vertices of the hypergraph. The
vertices that correspond to a bit-value $0$ are (supposedly) the
vertices of a Vertex Cover.

\medskip \noindent {\sc Vertices.} For each variable $x$ in layer $X_i$ we
construct a block of vertices $V[x]$. This block contains a vertex
for each subset of $R_{i}$. Throughout this section we slightly
abuse notation by writing a vertex rather than the set it
represents. The weight of the vertices inside the block $V[x]$ is
according to $\mu_p^{R_{i}}$, i.e. the weight of a subset $v
\subseteq R_{i}$ is proportional to $\mu_p^{R_{i}}(v) = p^{\card
v}(1-p)^{\card {R_{i}\setminus v}}$ as in
Definition~\ref{def:biasedmu}. All blocks in the same layer have
the same total weight and the total weight of each layer is
$\frac{1}{l}$. Formally, the weight of a vertex $v\in V[x]$ where
$x\in X_i$ is given by
 $$\frac{1}{l|X_i|} \mu_p^{R_{i}}(v).$$

\smallskip
\noindent {\sc Hyperedges.} We construct hyperedges between blocks
$V[x]$ and $V[y]$ such that there exists a constraint
$\pi_{x\rightarrow y}$. We connect a hyperedge between any
$v_1,\ldots,v_{k-1}\in V[x]$ and $u \in V[y]$ whenever
$\pi_{x\rightarrow y}(\bigcap_{i=1}^{k-1} v_i) \cap u = \phi$.

Let $IS(G)$ denote the weight of vertices contained in the largest
independent set of the hypergraph $G$.
\begin{lemma}[Completeness]
If $\Phi$ is satisfiable then $IS(G)\ge p$.
\end{lemma}
\begin{proof}
Let $A$ be a satisfying assignment for $\Phi$, i.e., $A$ maps each
$i\in [l]$ and $x\in X_i$ to an assignment in $R_{i}$ such that
all the constraints are satisfied. Let ${\cal I}\subseteq V$
contain in the block $V[x]$ all the vertices that contain the
assignment $A(x)$,
$$ {\cal I} = \bigcup_{x}\set{v\in V[x]}{v \ni A(x)} \,.$$

We claim that ${\cal I}$ is an independent set. Take any
$v_1,...,v_{k-1}$ in ${\cal I}\cap V[x]$ and a vertex $u$ in
${\cal I}\cap V[y]$. The vertices $v_1,\ldots,v_{k-1}$ intersect
on $A(x)$ and therefore the projection of their intersection
contains $\pi_{x\rightarrow y}(A(x))=A(y)$. Since $u$ is in ${\cal
I}\cap V[y]$ it must contain $A(y)$. The proof is completed by
noting that inside each block, the weight of the set of all
vertices that contain a specific assignment is exactly $p$.
\end{proof}

We now turn to the soundness of the construction.
\begin{lemma}[Soundness]
If $IS(G)\ge \epsilon$ then $\Phi$ is satisfiable.
\end{lemma}
This lemma completes the proof of our main result since the ratio
between the sizes of the vertex cover in the yes and no cases is
$\frac{1-\epsilon}{1-p}=(1-\epsilon)(k-1-\epsilon)$ which can be
arbitrarily close to $k-1$.
\begin{proof}
Let ${\cal I}$ be an independent set of weight $\epsilon$. We
consider the set $X'$ of all variables $x$ for which the weight of
${\cal I}\cap V[x]$ in $V[x]$ is at least $\epsilon/2$. A simple
averaging argument shows that the weight of $\bigcup_{x\in
X'}V[x]$ is at least $\frac{\epsilon}{2}$. Another averaging
argument shows that in at least
$\frac{\epsilon}{4}l=\frac{8}{\epsilon}$ layers, $X'$ contains at
least $\frac{\epsilon}{4}$ fraction of the variables. Using the
weak-density property of the PCP (see Definition~\ref{def:weak-density}), we conclude that there exist two
layers $X_i$ and $X_j$ such that $\frac{\epsilon^2}{64}$ fraction
of the constraints between them are constraints between variables
in $X'$. Let us denote by $X$ the variables in $X_i\cap X'$ and by
$Y$ the variables in $X_j\cap X'$.

For any variable $x\in X$, consider the vertices in ${\cal I} \cap
V[x]$. According to Lemma~\ref{intersecting_lemma} there exists a
$t=t(\frac{\epsilon}{2},k-1,p)$ and $k-1$ vertices in ${\cal
I}\cap V[x]$ that intersect in less than $t$ assignments. We
denote these vertices by $v_{x,1},\ldots,v_{x,k-1}$ and their
intersection by $B(x)$.

In the following we define an assignment to the variables in $X$
and $Y$ such that many of the constraints between them are
satisfied. Then Theorem \ref{pcp_thm} would imply that $\Phi$ must
be satisfiable (provided $R$ is chosen large enough). For a
variable $x\in X$ we choose a random assignment from the set
$B(x)$. For a variable $y\in Y$ we choose the assignment
$$A(y) = \mbox{maxvar}_{a\in R_Y} |\{ x\in X ~|~ a \in \pi_{x\rightarrow y}(B(x)) \}|,$$
i.e., the assignment that is contained in the largest number of
projections of $B(x)$.

Before continuing, we need the following simple claim:
\begin{claim}\label{popular_claim}
Let $A_1,\ldots,A_n$ be a collection of $n$ sets of size at most
$m$ such that no element is contained in more than $k$ sets. Then,
there are at least $\frac{n}{1+(k-1)m} \ge \frac{n}{km}$ disjoint
sets in this collection.
\end{claim}
\begin{proof}
We prove by induction on $n$ that there are at least
$\frac{n}{1+(k-1)m}$ disjoint sets in the collection. The claim
holds trivially for $n \le 1+(k-1)m$. Otherwise, consider all the
sets that intersect $A_1$. Since no element is contained in more
than $k$ sets, the number of such sets (including $A_1$) is at
most $1+(k-1)m$. Removing these sets we get, by using the
induction hypothesis, a collection that contains
$\frac{n-1-(k-1)m}{1+(k-1)m} = \frac{n}{1+(k-1)m} - 1$ disjoint
sets. We conclude the induction step by adding $A_1$ to the
disjoint sets.
\end{proof}

Consider a variable $y \in Y$ and a variable $x$ such that the
constraint $\pi_{x \rightarrow y}$ exists. There are no hyperedges of
the form $(v_{x,1},\ldots,v_{x,k-1},u)$ for any vertex $u\in {\cal
I}\cap V[y]$. Therefore, every  vertex $u \in {\cal I}\cap V[y]$
must intersect  $\pi_{x \rightarrow y}(B(x))$. Now consider the
family of projections $\pi_{x \rightarrow y}(B(x))$ for all the
variables $x$ such that the constraint $\pi_{x \rightarrow y}$ exists.
Let $q$ denote the maximum number of disjoint sets inside this
family. Note that every disjoint set reduces the weight of the
vertices in ${\cal I}\cap V[y]$ by a factor of $1-(1-p)^t$.
Because the weight of ${\cal I}\cap V[y]$ is at least
$\frac{\epsilon}{4}$, we obtain that $q$ is at most
$\log(\frac{\epsilon}{4})/\log(1-(1-p)^{t})$.
Claim~\ref{popular_claim} implies that there exists an assignment
for $y$ that is contained in at least a fraction
\[ \frac{1}{t\log(\frac{\epsilon}{4})/\log(1-(1-p)^{t})} \]
of the projections $\pi_{x \rightarrow y}(B(x))$. Therefore, the
expected fraction of constraints satisfied between $X$ and $Y$ is at
least
\[ \frac{1}{t^2\log(\frac{\epsilon}{4})/\log(1-(1-p)^{t})}\]
which is a constant that does not depend on $R$. We complete the
proof by choosing the range $R$ of the PCP large enough so that
this fraction is larger than $1/R^\gamma$ and applying Theorem
\ref{pcp_thm}. This completes the soundness proof.
\end{proof}

\end{proof}

\section{Acknowledgements}
We would like to thank Noga Alon for his help with $s$-wise
$t$-intersecting families.


\end{document}